\documentclass[prd,preprint,tightenlines,floatfix,showpacs,preprintnumbers,nofootinbib,eqsecnum]{revtex4}

 \usepackage[dvips,final]{graphicx}
  \usepackage{amssymb}
   \usepackage{amsmath} % Advanced math typesetting
    \usepackage{amsfonts}
     \usepackage{epsfig}
     \usepackage{color}
      \usepackage{bm} % bold math
        \usepackage{xcolor}
        
\usepackage{booktabs}
\usepackage{multirow}
\usepackage{slashed}
\usepackage{comment}

\def\doublespace{\def\baselinestretch{1.6}\large\normalsize}
\def\normalspace{\def\baselinestretch{1.0}\normalsize}
\def\PSfig#1#2{\scalebox{#1}{\includegraphics{#2}}}

\def\Caption#1{
  \normalspace
  \vskip-1mm\caption{\sl#1}\vskip-1mm
  \doublespace
}

\def\BA{\begin{eqnarray}}
\def\BE{\begin{equation}}
\def\BF{\begin{figure}[htb]}
\def\BT{\begin{table}[htb]}
\def\EA{\end{eqnarray}}
\def\EE{\end{equation}}
\def\EF{\end{figure}}
\def\ET{\end{table}}

\def\ra{\rangle}

\def\mb{\,\mbox{mb}}

\def\TeV{\,\mbox{TeV}}
\def\GeV{\,\mbox{GeV}}

\def\Jpsi{J\!/\!\psi}
\def\psip{\psi^{\,\prime}}
\def\psipp{\psi^{\,\prime\prime}}
\def\Y{\Upsilon}
\def\Yp{\Upsilon^{\,\prime}}
\def\Ypp{\Upsilon^{\,\prime\prime}}
\def\sqq{\sigma_{Q\bar Q}}

\def\aem{\alpha_{em}}

\def\lsim{\mathrel{\rlap{\lower4pt\hbox{\hskip1pt$\sim$}}
     \raise1pt\hbox{$<$}}}         %less than or approx. symbol
 \def\gsim{\mathrel{\rlap{\lower4pt\hbox{\hskip1pt$\sim$}}
     \raise1pt\hbox{$>$}}}         %greater than or approx. symbol
\begin{document}
%%%%%%%%%%%%%%%%%

%================================================
\title{
%\vspace*{-2.0cm}
D-wave effects in
heavy quarkonium production \\
in ultraperipheral nuclear collisions
}
%================================================

\author{M. Krelina$^{1}$}
\email{michal.krelina@cvut.cz}

\author{J. Nemchik$^{1,2}$}
\email{nemcik@saske.sk}

%\vspace*{0.5cm}

\affiliation{$^1$
Czech Technical University in Prague, FNSPE, B\v rehov\'a 7, 11519
Prague, Czech Republic}
\affiliation{$^2$
Institute of Experimental Physics SAS, Watsonova 47, 04001 Ko\v sice, Slovakia
}

\date{\today}
%%%%%%%%%%%%%%%%%%%%%%%%%
\begin{abstract}
%%%%%%%%%%%%%%%%%%%%%%%%%
\vspace*{5mm}

The $D$-wave admixture in quarkonium wave functions is acquired from the photonlike structure of $V\to Q\bar Q$ transition in the light-front frame widely used in the literature. Such a $D$-wave ballast is not justified by any nonrelativistic model for $Q-\bar Q$ interaction potential and leads to falsified predictions for the  cross sections in  heavy quarkonium production in ultraperipheral nuclear collisions. We analyze this negative role of $D$-wave contribution by comparing with our previous studies based on a simple non-photon-like ``$S$-wave-only'' $V\to Q\bar Q$ transition in the $Q\bar Q$ rest frame. 

%%%%%%%%%%%%%%%%%%%%%%%%
\end{abstract}
%%%%%%%%%%%%%%%%%%%%%%%%


\maketitle

%%%%%%%%%%%%%%%%%%%%%%%%%
\section{Introduction}
%%%%%%%%%%%%%%%%%%%%%%%%%

Recent investigations of heavy quarkonium [$V=\Jpsi(1S)$, $\psip(2S)$, $\psipp(3S)$, ..., $\Y(1S)$, $\Yp(2S)$,  $\Ypp(3S)$, ...] production in ultraperipheral collisions (UPC) at the Relativistic Heavy Ion Collider (RHIC) and the Large Hadron Collider (LHC) is very effective
for theoretical study of various nuclear effects occurring in diffractive photoproduction off nuclei.
Although 
our investigation of the corresponding photoproduction mechanism on the proton target
within the light-front (LF) color dipole formalism  has a long-standing tradition 
\cite{Kopeliovich:1991pu,Kopeliovich:1993pw,Nemchik:1996pp,Nemchik:1996cw,Nemchik:2002ug,Krelina:2018hmt,Cepila:2019skb,Krelina:2019egg} there are still open questions associated mainly with the structure of the $V\to Q\bar Q$ transition.
The most
phenomenological studies of diffractive electroproduction of heavy quarkonia are based on an unjustified assumption of a similar structure for $\gamma^*\to Q\bar Q$ and $V\to Q\bar Q$ vertices. This leads to an extra $D$-wave admixture in the photonlike $V\to Q\bar Q$ vertex in the $Q\bar Q$ rest frame.  However, any realistic nonrelativistic $Q\bar Q$ potential model cannot prove the relative weight of such a spurious $D$-wave component contribution. For this reason, in the present paper, we analyze its magnitude using, besides the standard photonlike structure in the LF frame, also our previous studies
\cite{Kopeliovich:2020has} based on the ``$S$-wave-only'' $V\to Q\bar Q$ transition.
 
In Ref.~\cite{Krelina:2019egg} we have studied the relative contribution of $D$-wave component in diffractive electroproduction of heavy quarkonia off proton targets. Here we compared the both structures of $V\to Q\bar Q$ transitions, the standard photonlike structure in the LF frame with a simple  ``$S$-wave-only'' structure in the $Q\bar Q$ rest frame \cite{Krelina:2018hmt,Cepila:2019skb}. We have found that for $1S$ charmonium photoproduction the relative undesirable impact of the $D$-wave component on the magnitude of production cross section is not large and represents about $5\%-10\%$ depending on the photon energy. Not so for radially excited charmonia where the nodal structure of their wave functions leads to a boosting of the negative role of $D$-wave admixture in estimations of production cross sections causing their $20\%-30\%$ enhancement.

In the present paper, we extend such a study for heavy quarkonium production in UPC analyzing thus for the first time the relative contribution of undesirable admixture of $D$-wave component to nuclear cross sections as a function of rapidity and collision energy
$\sqrt{s_N}$. Treating the UPC, besides $D$-wave effects, the onset of another nuclear phenomena can affect the quarkonium production rate as was analyzed in Ref.~\cite{Kopeliovich:2020has} within the LF QCD dipole formalism.
They concern the higher twist effect related to the lowest $Q\bar Q$ Fock component of the photon, as well as the leading twist effect associated with higher photon components containing gluons.

The former effect represents the quark shadowing controlled by the distance called the {\it coherence length} (CL) \cite{Kopeliovich:1991pu,Kopeliovich:2001xj}, which can be expressed in the rest frame of the nucleus as,
%
%----------------------------------------
 \BE
l_c = \frac{2\,q}{m_V^2} = \frac{s - m_N^2}{m_N\,m_V^2}\ ,
\label{lc}
 \EE
%----------------------------------------
%
where $q$ is the photon energy and
$m_N$ and $m_V$ are the nucleon and quarkonium mass, respectively.
Following results from our recent paper ~\cite{Kopeliovich:2020has}, this phenomenon
has been incorporated via the finite-$l_c$ 
correction factors calculated within the rigorous Green function formalism, which naturally includes the CL effects.

The latter effect is known as the {\it gluon shadowing} (GS) and is treated in terms of the LF QCD dipole approach. The corresponding CL of multigluon Fock state is shorter \cite{Kopeliovich:2000ra} compared to the
lowest $Q\bar Q$ photon fluctuations and so the higher photon energy is required for manifestation of the corresponding shadowing correction. Similarly as for quarks, in the present paper, we calculate the GS correction adopting the Green function formalism \cite{Kopeliovich:1999am} with improvements from Ref.~\cite{Krelina:2020ipn}.

Besides the above shadowing corrections, the nuclear suppression is affected also by
the final state absorption of produced quarkonia
\cite{Kopeliovich:1991pu,Kopeliovich:1993gk,Kopeliovich:1993pw,Nemchik:1994fp,Nemchik:1994fq,Nemchik:1996cw,Hufner:2000jb,jan-00a,jan-00b,Kopeliovich:2001xj,Kopeliovich:2007wx}.
It is related to
the phenomenon known as the {\it color transparency}, where the
photon fluctuations with a smaller transverse size associated with a larger quark mass are less absorbed during propagation through the medium. The corresponding evolution of the small-sized $Q\bar Q$ photon component to the normal-sized quarkonia is controlled 
by the length scale known as the {\it formation length}.
In the rest frame of the nucleus it has the following form \cite{Kopeliovich:1991pu,Kopeliovich:2001xj},
%
%-------------------------------------------
 \BE
l_f = \frac{2\,q}
{\left.m_{V^\prime}\right.^2 - \left.m_V\right.^2}\ ,
\label{tf}
 \EE
%--------------------------------------------
%
where $m_{V^\prime}$ is the quarkonium mass in the $2S$ state.

In our analysis of a negative role of the $D$-wave component in quarkonium wave functions, we treat only the production of $S$-wave quarkonia since their wave functions can be simply factorized into radial and spin-dependent parts. The former part can be acquired properly
in the $Q\bar Q$ rest frame as a solution of the Schr\"odinger equation for various realistic $Q-\bar Q$ interaction potentials proposed in the literature. As an example, in the present paper we adopt the Buchm\"uller-Tye (BT) \cite{Buchmuller:1980su} and the power-like (POW) \cite{Martin:1980jx,Barik:1980ai} potential. The choice of other potentials has practically no impact on the magnitude of analyzed 
$D$-wave effects.
The same argument concerns our preference to adopt the Kopeliovich-Schafer-Tarasov
(KST) ~\cite{Kopeliovich:1999am} and Golec-Biernat–Wusthoff (GBW) ~\cite{GolecBiernat:1998js,GolecBiernat:1999qd} models
for the dipole cross section.

The photonlike structure of the quarkonium vertex is treated directly in the LF frame,
what requires only the Lorentz boost of radial components of quarkonium wave functions 
from the $Q\bar Q$ rest frame.
Here we adopt a widely used procedure known as the Terentev prescription \cite{Terentev:1976jk}. However, a simple ``$S$-wave-only'' $V\to Q\bar Q$ structure in the $Q\bar Q$ rest frame requires to perform additionally the corresponding boost also for the spin-dependent components known as the {\it Melosh spin rotation} \cite{Melosh:1974cu,Hufner:2000jb,Krelina:2018hmt,Cepila:2019skb,Krelina:2019egg}. 

The paper is organized as follows. In the next section
we present basic expressions for calculation of nuclear cross sections,
separately for coherent (elastic), as well as incoherent (quasielastic) heavy quarkonium production in UPC.
Section~\ref{Sec:2} is devoted to the analysis of the undesirable $D$-wave admixture,
related to the photonlike structure of $V\to Q\bar Q$ transition, together with estimations of the corresponding impact on magnitudes of nuclear cross sections.
Finally, Sec.~\ref{Sec:3} contains a summary with the main concluding remarks on
how the negative role of $D$-wave component can be 
identified by the future measurements.

%%%%%%%%%%%%%%%%%%%%%%%%%%%%%%%%%%%%%%%%%%%%%%%%%%%%%%
\section{Basic formulas in the color-dipole formalism}
\label{Sec:1}
%%%%%%%%%%%%%%%%%%%%%%%%%%%%%%%%%%%%%%%%%%%%%%%%%%%%%%

Within the one-photon-exchange approximation
in the rest frame of the target nucleus $A$,
the cross section for the
photoproduction of a vector meson $V$ by the Weizs\"acker-Williams photons
reads
%-------------------------------------------------------------------
\BE
  q\frac{d\sigma}{dq} = \int\,d^2\tau \int\,d^2b \,\,
  n(q,\vec b-\vec\tau)\, 
  \frac{d^2
  \sigma_A(s,b)}{d^2b}\,,
\label{cs-upc}
\EE
%------------------------------------------------------------------
where $\vec\tau$ is the relative impact parameter of a nuclear collision, $\vec{b}$ is the impact parameter of the photon-nucleon collision
relative to the center of one of the nuclei and the variable
$n(q,\vec b)$ represents the photon flux induced by the
projectile nucleus, 
%-----------------------------------------------------------------
\BE
  n(q,\vec b) = \frac{\aem Z^2 q^2}{\pi^2\gamma^2}\,
%  \Biggl [
  K_1^2\left(\frac{b\,q}{\gamma}\right)
%  +
%  \frac{1}{\gamma^2} K_0^2\left(\frac{bk}{\gamma}\right)
%  \Biggr ]
  \,,
  \label{flux}
\EE
%----------------------------------------------------------------
where $Z$ is the ion charge, $\aem= 1/137$ is the fine-structure constant,
$K_{1}$ is the modified Bessel function and the Lorentz factor $\gamma = s_N/2 m_N^2$.

\vspace*{0.1cm}
%%%%%%%%%%%%%%%%%%%%%%%%%%%%%%%%%%%%%%%%%%%%%%%%%
\textbf{Coherent production (coh).}
%%%%%%%%%%%%%%%%%%%%%%%%%%%%%%%%%%%%%%%%%%%%%%%%%
In the LF dipole approach (see Refs.~\cite{Kopeliovich:1991pu,Hufner:2000jb,Krelina:2018hmt,Cepila:2019skb,Ivanov:2002kc,Nemchik:2002ug}, for example), assuming large photon energies when $l_c\gg R_A$, where $R_A$ is the nuclear radius, the corresponding coherent cross section for the process 
$\gamma A\to V\!A$ (the nucleus remains intact) takes a simple asymptotic form,
%-----------------------------------------------------------------
 \BA
 \frac{d^2\sigma_A^{coh}(s,b)}{d^2b}
\Biggr|_{l_c \gg R_A} 
&=&
\Biggl |
\int d^2r\int_0^1 d\alpha\,
\Psi_{V}^{*}(\vec r,\alpha)\,
\Sigma_{A}^{coh}(r,s,b)\,
\Psi_{\gamma}(\vec r,\alpha)
\Biggr |^2,
\nonumber\\
\Sigma_{A}^{coh}(r,s,b)\,
&=&
1 -
\exp\left[-\frac{1}{2}\,\sqq(r,s)\,T_A(b)\right]
\label{coh-lcl}.
 \EA
%-------------------------------------------------------------------
%
Here 
$\Psi_V(r,\alpha)$ is the LF wave function for heavy quarkonium and
$\Psi_{\gamma}(r,\alpha)$ is the LF distribution 
of the $Q\bar Q$ Fock component of the quasireal photon,
where the $Q\bar Q$ fluctuation (dipole) has the transverse size $\vec{r}$ and the variable $\alpha = p_Q^+/p_{\gamma}^+$ is the boost-invariant fraction of the photon momentum carried by a heavy quark (or antiquark). The variable
$T_A(b) = \int_{-\infty}^{\infty} dz\,\rho_A(b,z)$ represents the nuclear thickness function normalized as $\int d^2 b\,T_A(b) = A$, 
where $\rho_A(b,z)$ is the nuclear density function of realistic Wood-Saxon form with parameters taken from \cite{DeJager:1987qc}.
The universal dipole-nucleon total cross section $\sqq(r,s)$, introduced in Ref.~\cite{zkl}, depends on the transverse $Q-\bar Q$ separation $r$ and c.m. energy squared $s = m_V\,\sqrt{s_{N}}\,\exp[y]$, respectively
variable $x = m_V^2/W^2 = m_V\,\exp{[-y]}/\sqrt{s_{N}}$, where
$y$ is the rapidity. 

\vspace*{0.1cm}
%%%%%%%%%%%%%%%%%%%%%%%%%%%%%%%%%%%%%%%%%%%%%%%%%%%
\textbf{Incoherent production (inc).}
%%%%%%%%%%%%%%%%%%%%%%%%%%%%%%%%%%%%%%%%%%%%%%%%%%%
Here the vector meson is produced in a quasielastic process $\gamma A\to V A^*$, where the nucleus is in the excited state. In the high energy limit, $l_c\gg R_A$, the corresponding nuclear cross section reads \cite{Kopeliovich:2020has},
%
%--------------------------------------------------------------------------------
\BA 
\frac{d^2\sigma_A^{inc}(s,b)}{d^2b}
\Biggr|_{l_c \gg R_A} 
&\approx&
\frac{T_A(b)}{16 \pi B(s)}\,
\Biggl |
\int d^2r\int_0^1 d\alpha\,
\Psi_{V}^{*}(\vec r,\alpha)\,
\Sigma_{A}^{inc}(r,s,b)\,
\Psi_{\gamma}(\vec r,\alpha)\,
\Biggr |^2
\nonumber\\
\Sigma_{A}^{inc}(r,s,b)\,
&=&
\sqq(r,s)\,
\exp\left[-\frac{1}{2} \sqq(r,s)\, T_A(b)\right]\,,
\label{inc-lcl}
 \EA
%---------------------------------------------------------------------------------
%
where
$B(s)$ is the slope parameter in reaction $\gamma N\to V N$.

\vspace*{0.1cm}
%%%%%%%%%%%%%%%%%%%%%%%%%%%%%%%%%%
\textbf{Scenario I.}
%%%%%%%%%%%%%%%%%%%%%%%%%%%%%%%%%%
In the conventional standard and frequently used 
\textit{scenario I}, corresponding to the photonlike $V\to Q\bar Q$ transition directly in the LF frame without the Melosh transform, the imaginary part of the 
$\gamma N\to V N$ amplitude has the following structure \cite{Krelina:2019egg},
%-----------------------------------------------------
\BA
&& \mathrm{Im}\mathcal{A}_1(s)
=
N_1\,
\int_0^1 d\alpha \int d^2r \,
\sqq(r,s)
\left[\Sigma^{(1)}_{}(r,\alpha) 
+
\Sigma^{(2)}_{}(r,\alpha)\right]\,,   
 \nonumber
\\
&&\Sigma^{(1)}_{}(r,\alpha) 
= 
m_Q^2\,
K_0(m_Q r)\,
\int_0^\infty d p_T\,p_T\,J_0(p_T r)\Psi_{V}(\alpha,p_T)\,, 
\nonumber \\
&&\Sigma^{(2)}_{}(r,\alpha) 
= 
m_Q\, \bigl [\alpha^2 + (1-\alpha)^2\bigr ]\,
K_1(m_Q r)\,
\int_0^\infty d p_T\,p_T^2\,J_1(p_T r)\Psi_{V}(\alpha,p_T)
\,.
%.........
\label{AT1}
%.........  
\EA
%------------------------------------------------------
%
Here $N_1=Z_Q\,\sqrt{2 N_c^2 \,\alpha_{em}}/2\,\pi$, the factor $N_c=3$ represents the number of colors in QCD, $Z_Q$ is
the electric charge of the heavy quark,
$J_{0,1}$ and $K_{0,1}$ are the Bessel functions of the first kind and the modified Bessel functions of the second kind, respectively.

\vspace*{0.1cm}
%%%%%%%%%%%%%%%%%%%%%%%%%%%%%%%%%%%
\textbf{Scenario II.}
%%%%%%%%%%%%%%%%%%%%%%%%%%%%%%%%%%%
For a simple ``$S$-wave-only'' structure of the $V\to Q\bar Q$ vertex in the $Q\bar Q$ rest frame (\textit{scenario II}) 
\cite{Ivanov:2002kc,Ivanov:2002eq,Ivanov:2007ms,Krelina:2018hmt,Cepila:2019skb},
with subsequent Melosh spin transformation to the LF frame,
the corresponding photoproduction amplitude reads,
%
%------------------------------------------------------
\BA
&& 
\!\!\!\!\!\!\!\!\!\!
\mathrm{Im}\mathcal{A}_2(s)
=
\frac{N_2}{2}\,
\int_0^1 d\alpha \int d^2r \,
\sqq(r,s)
\left[\Sigma^{(1)}_{M}(r,\alpha) 
+
\Sigma^{(2)}_{M}(r,\alpha)\right]\,,
 \nonumber
\\ 
&& 
\!\!\!\!\!\!\!\!\!\!
   \Sigma^{(1)}_M(r,\alpha)
   =  
   %\frac{Z_Q\,\sqrt{N_c \,\alpha_{em}}}{2\,\pi\,\sqrt{2}}\,
   K_0(m_Q r) \int_0^\infty dp_T\,p_T\,
%   \nonumber\\
%   &\times&
   J_0(p_T r) \Psi_V (\alpha,p_T) 
   \left[ \frac{2\,m_Q^2(m_L+m_T)+m_L\,p_T^2}{ m_T (m_L + m_T)} \right]\,,
   \label{sigma-sr1}
\nonumber \\
&&
\!\!\!\!\!\!\!\!\!\!
  \Sigma^{(2)}_M(r,\alpha)
   = 
   %\frac{Z_Q\,\sqrt{N_c \,\alpha_{em}}}{2\,\pi\,\sqrt{2}}\,
   K_1(m_Q r) \int_0^\infty dp_T\,p_T^2\,
%   \nonumber\\
 %  &\times&   
   J_1(p_T r) \Psi_{V} (\alpha,p_T) \left[
%   \frac{p_T}{2}\,  
   \frac{m_Q^2(m_L+2m_T)-m_T\,m_L^2}{m_Q\,m_T (m_L+m_T)} \right]\,,
   \label{sigma-sr2}
\EA
%-------------------------------------------------------
%
where $N_2=Z_Q\,\sqrt{2 N_c \,\alpha_{em}}/2\,\pi$,
%==============================
$m_T = \sqrt{m_Q^2 + p_T^2}$
%==============================
and 
%========================================
$m_L = 2\, m_Q\,\sqrt{\alpha(1-\alpha)}$\,.
%========================================

Our model calculations include also a small real part 
\cite{bronzan-74,Nemchik:1996cw,forshaw-03} of the $\gamma N\to V N$ amplitude
 performing the following replacement in Eqs.~(\ref{AT1}) and (\ref{sigma-sr2}),
%
%------------------------------------------------------
\BA
\sqq(s,r)
\Rightarrow
\sqq(s,r)
\,\cdot
\left(1 - i\,\frac{\pi}{2}\,\frac
{\partial
 \,\ln\,{\sqq(s,r)}}
{\partial\,\ln s} \right)\ .
%--------------
  \label{re/im}
%--------------
\EA
%-----------------------------------------------------
%
The expression (\ref{AT1}) corresponding to \textit{scenario I} and the expression (\ref{sigma-sr2}) related to \textit{scenario~II} can be straightforwardly generalized to nuclear targets via replacements $\sqq\Rightarrow\Sigma_{A}^{coh}$ and $\sqq\Rightarrow\Sigma_{A}^{inc}$,
where 
$\Sigma_{A}^{coh}$ and $\Sigma_{A}^{inc}$ are determined by Eqs.~(\ref{coh-lcl}) and (\ref{inc-lcl}).

%
%====================================================================================
Expressions~(\ref{coh-lcl}) and (\ref{inc-lcl}) 
lead to quark shadowing correction which is a higher twist effect related to the size of the $Q\bar Q$ photon fluctuations and vanishes as $1/m_c^2$
and $1/m_b^2$ in photoproduction of charmonia and bottomonia, respectively. 
Although Eqs.~(\ref{coh-lcl}) and (\ref{inc-lcl}) allow to calculate nuclear cross sections only
at large photon energies corresponding to long CL, $l_c\gg R_A$,
in our analysis we have also included finite-$l_c$ corrections which have been calculated using a rigorous Green function formalism as presented in Ref.~\cite{Kopeliovich:2020has}.

Another nuclear phenomenon incorporated in our study of the negative role of the $D$-wave component is associated with leading twist gluon shadowing related to higher Fock components of the photon containing gluons, i.e. $|Q\bar QG\ra$, $|Q\bar Q2G\ra$ ... $|Q\bar QnG\ra$. 
Here for calculations of the gluon shadowing factor $R_G$,
defined as the nucleus-to-nucleon ratio of gluon densities, $R_G(s,\mu^2,b) = G_A(s,\mu^2,b)/(A\,G_N(s,\mu^2))$,
we have adopted the path integral technique as well with details as described in Refs.~\cite{Krelina:2020ipn,Kopeliovich:2020has}.
At small dipole sizes $\vec r$
the dipole cross section $\sqq(r,s)$ depends on the gluon distribution in the target.
Then the nuclear shadowing of the gluon distribution should reduce $\sqq(r,s)$ in nuclear reactions with respect to processes on proton targets.
Consequently, the GS corrections are included in our calculations via the following substitution in Eqs.~(\ref{coh-lcl}) and (\ref{inc-lcl}),
%
%------------------------------------------------------------
\BE
  \sqq(r,s) \Rightarrow \sqq(r,s) \cdot R_G(s,\mu^2,b)\,,
\label{eq:dipole:gs:replace}
\EE
%------------------------------------------------------------
%
where the GS factor $R_G$
is probed at the factorization scale $\mu^2\sim\tau M_V^2$ with $\tau = B/Y^2$. 
Here the scale factor $B\approx 9 - 10$ ~\cite{Nikolaev:1994ce} and
factors $Y\sim 5 - 6$ have been estimated in Ref.~\cite{Krelina:2018hmt} for various quarkonium states.
Finally, we would like to emphasize that both corrections, the quark and gluon shadowing, are
proportional to the gluon density in the target and steeply rise with energy.

%===========================================================================================

%%%%%%%%%%%%%%%%%%%%%%%%%%%%%%%%%%%%%%%%%%%%%%%%%%%%%%%%%%%%%%%%%%%%
\BF
\hspace*{-0.3cm}
\PSfig{1.1}{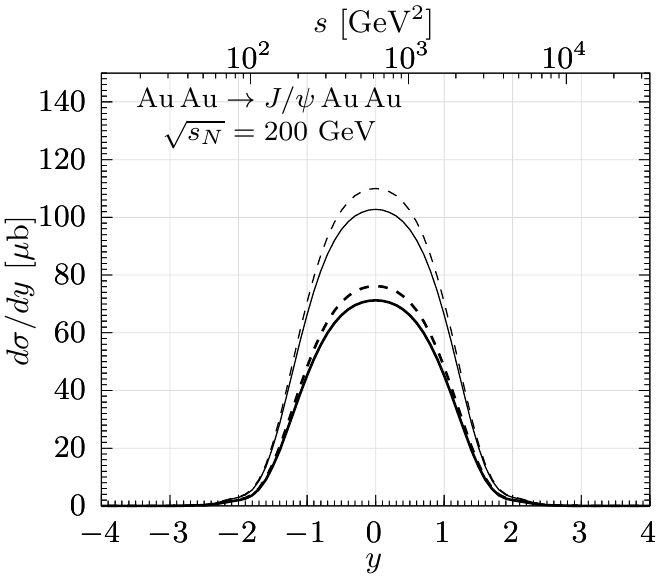}~~~~
\PSfig{1.1}{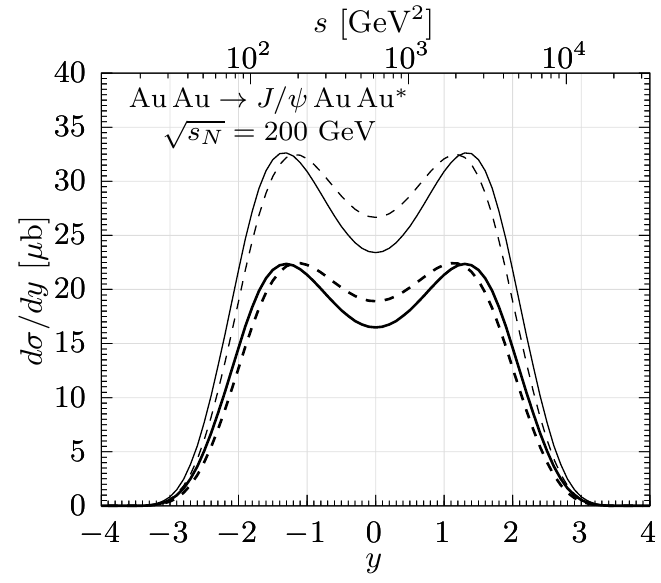}~~~~\\
\hspace*{-0.4cm}
\PSfig{1.1}{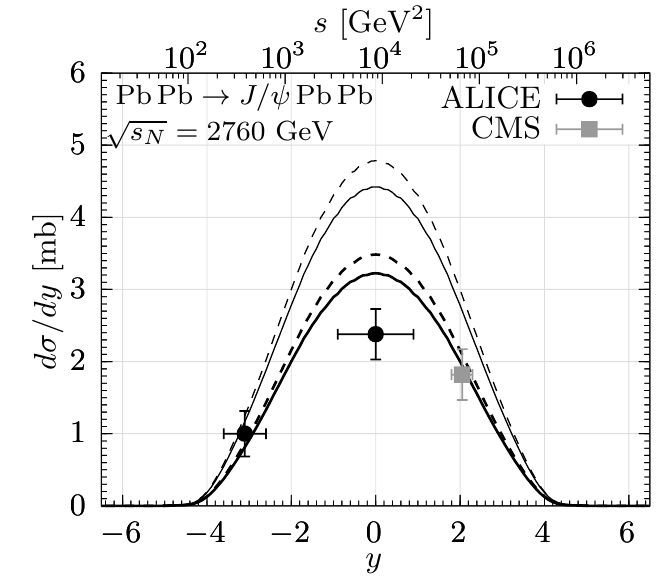}~~~~
\PSfig{1.1}{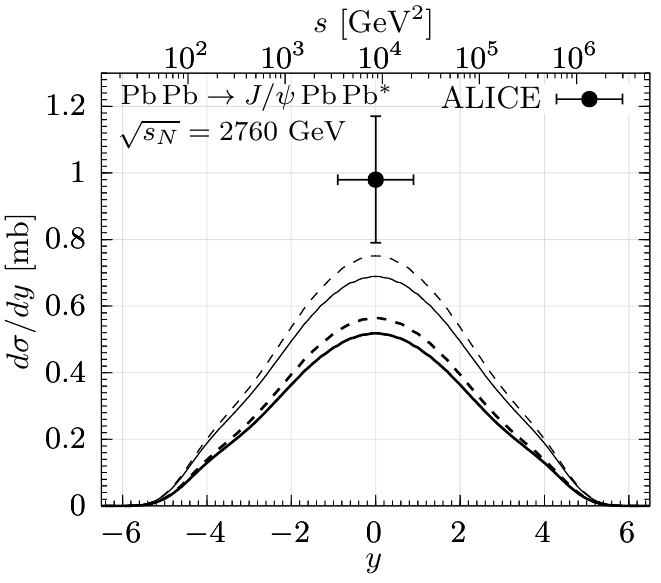}~~~~\\
\PSfig{1.1}{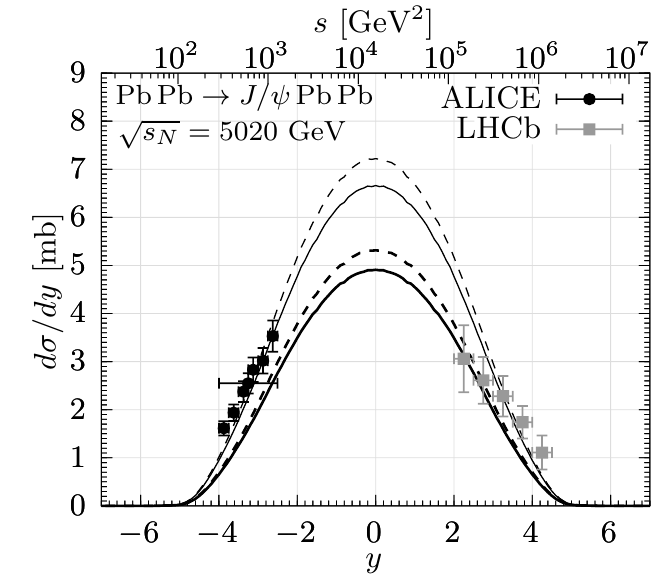}~~~~
\PSfig{1.1}{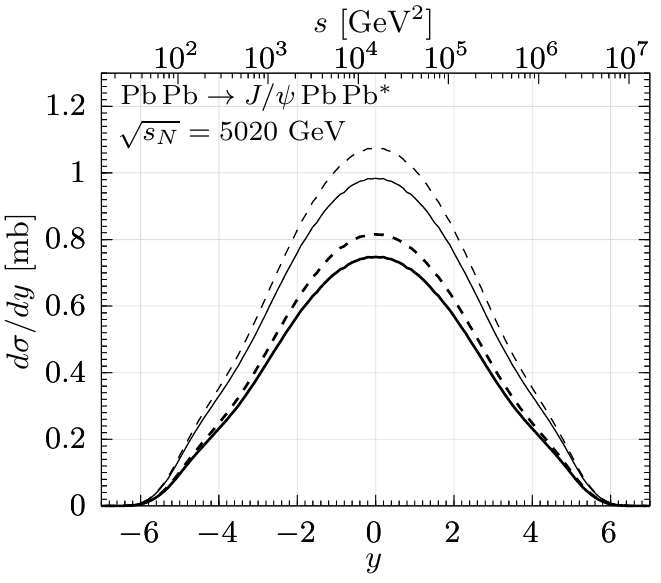}~~~~
\vspace*{-0.30cm}
\Caption{
  \label{Fig-UPC1psi}
  Manifestation of $D$-wave component in
  rapidity distributions of coherent (left panels) and incoherent (right panels) charmonium
  photoproduction in UPC at RHIC collision energy $\sqrt{s_N}=200\,\GeV$ (top panels) and at LHC energies $\sqrt{s_N} = 2.76\,\TeV$ (middle panels) and $\sqrt{s_N} = 5.02\,\TeV$ (bottom panels). The nuclear cross sections are calculated with charmonium wave functions generated by the POW (thin lines) and BT (thick lines) potential and with the GBW model for the dipole cross section. The dashed and solid lines correspond to photonlike $\Jpsi\to c\bar c$ transition in the LF frame and to a simple "S-wave only" charmonium vertex in the $c\bar c$ rest frame, respectively. 
  Model predictions are compared with 
  data from 
  CMS \cite{Khachatryan:2016qhq}, ALICE \cite{Abelev:2012ba,Abbas:2013oua,Adam:2015sia,Acharya:2019vlb} and  LHCb \cite{LHCb:2018ofh} collaborations.
  }
\EF
%%%%%%%%%%%%%%%%%%%%%%%%%%%%%%%%%%%%%%%%%%%%%%%%%%%%%%%%%%%%%%%%%%%%

%%%%%%%%%%%%%%%%%%%%%%%%%%%%%%%%%%%%%%%%%%%%%%%%%%%%%%%%%%%%%%%%%%%%%
\section{Analysis of the relative $D$-wave contribution}
\label{Sec:2}
%%%%%%%%%%%%%%%%%%%%%%%%%%%%%%%%%%%%%%%%%%%%%%%%%%%%%%%%%%%%%%%%%%%%%

For calculation of the incoherent nuclear cross sections (\ref{inc-lcl}) 
we rely on the standard Regge form for the slope parameter,
%=====================================================
$B_{\Jpsi}(s)=B_0 + 2\,\alpha'(0) \ln\big(s/s_0\big)$,
%=====================================================
with the parameters $\alpha'=0.171\GeV^{-2}$, $B_{0}=1.54\,\GeV^{-2}$ and 
$s_0=1\GeV^2$ fitted in \cite{Cepila:2019skb}. For $1S$-bottomonium photoproduction
we used values of $B_{\Y}(s)\approx B_{\Jpsi}(s) -1\GeV^{-2}$ \cite{Cepila:2019skb}.
Here we have also found that a very weak node effect in photoproduction of the $\Yp(2S)$ state causes a similarity $B_{\Yp}(s)\sim B_{\Y}(s)$. However, for production of $\psip(2S)$ one has to include the difference in diffraction slopes $\Delta_B(s)=B_{\Jpsi}(s)-B_{\psip}(s)$
with the parametrization of the factor $\Delta_B(s)$ from Ref.~\cite{Cepila:2019skb} (see also Ref.~\cite{jan-98}).

In our calculations, we consider that the photonuclear reaction can be also induced
by the photon from the second nucleus of the colliding nuclei in UPC 
%performing 
%implementing
via replacement in Eq.~(\ref{cs-upc}) $y\Rightarrow -y$. 
The KST and GBW phenomenological models for the dipole cross section used in our analysis 
are poorly known at small $s$ corresponding to large values of $x = m_V^2/s$. Here we
rely on dipole model modification 
by an additional factor $(1-x)^7$ ~\cite{Kutak:2003bd}.

Because of a weak sensitivity of charmonium results to a choice of the model for the dipole cross section $\sqq$, we analyzed a negative role of the $D$-wave component in quarkonium wave functions adopting two distinct realistic $Q-\bar Q$ interaction potentials, 
powerlike (POW) and Buch\"uller-Tye (BT).
These potentials generate the quarkonium wave functions in the $Q\bar Q$ rest frame since they are inherent in the corresponding Schr\"odinger equation.
On the other hand, the production of bottomonium states is more sensitive to the choice of the model for $\sqq$ as was studied in Ref.~\cite{Cepila:2019skb}. For this reason, the investigation of the corresponding $D$-wave effects has been realized adopting two distinct models for $\sqq$ - KST and GBW.
In both cases the study of undesirable $D$-wave contributions to magnitudes of nuclear cross sections has been performed including nuclear phenomena, such as the gluon shadowing and the finite-$l_c$ corrections described in detail in Ref.~\cite{Kopeliovich:2020has}.

Figure~\ref{Fig-UPC1psi} shows 
our predictions for the rapidity distributions of coherent (left panels) and incoherent (right panels) charmonium photoproduction in UPC vs. the LHC data from the CMS \cite{Khachatryan:2016qhq} and ALICE \cite{Abelev:2012ba,Abbas:2013oua,Adam:2015sia} collaborations at c.m. collision energy $\sqrt{s_N} = 2.76\,\TeV$, as well as the LHCb \cite{LHCb:2018ofh} and ALICE \cite{Acharya:2019vlb} data at $\sqrt{s_N} = 5.02\,\TeV$.
The corresponding calculations have been performed 
at $\sqrt{s_N} = 200\,\GeV$ (top panels),
$\sqrt{s_N} = 2.76\,\TeV$ (middle panels) and $\sqrt{s_N} = 5.02\,\TeV$ (bottom panels)
adopting the GBW model for the dipole cross section.
Here charmonium wave functions are generated by the POW (thin lines) and BT (thick lines)
potential.
\textit{Scenarios I} and \textit{II}, corresponding to the photonlike quarkonium vertex with $D$-wave admixture and ``$S$-wave-only'' $V\to Q\bar Q$ transition, are depicted by dashed and solid lines, respectively.

%%%%%%%%%%%%%%%%%%%%%%%%%%%%%%%%%%%%%%%%%%%%%%%%%%%%%%%%%%%%%%%%%%%%
\BF
\PSfig{1.1}{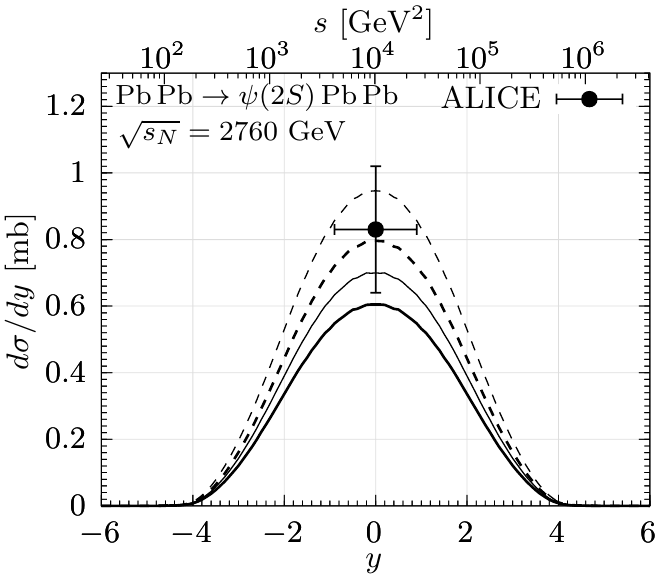}~~~~
\PSfig{1.1}{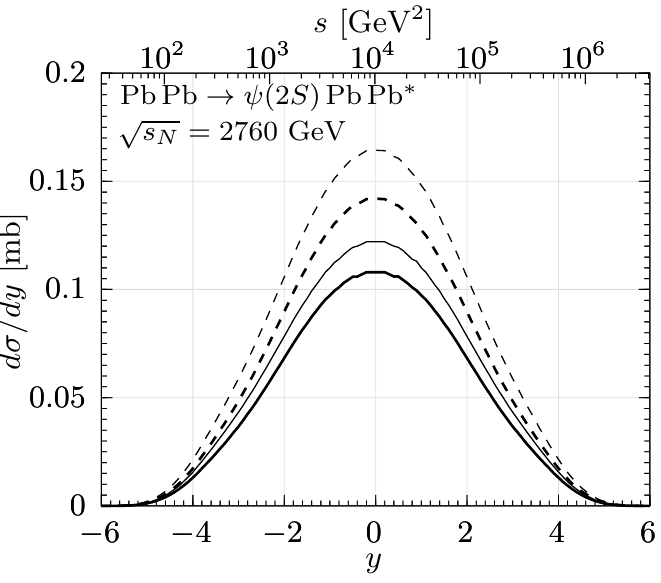}~~~~\\
\PSfig{1.1}{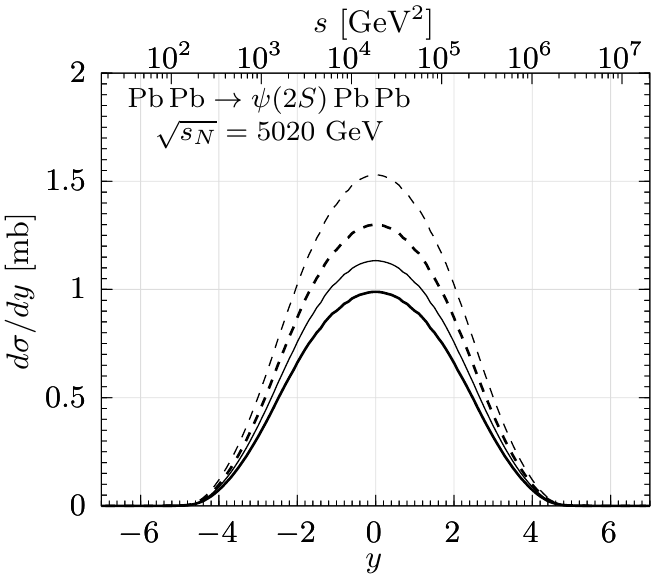}~~~~
\PSfig{1.1}{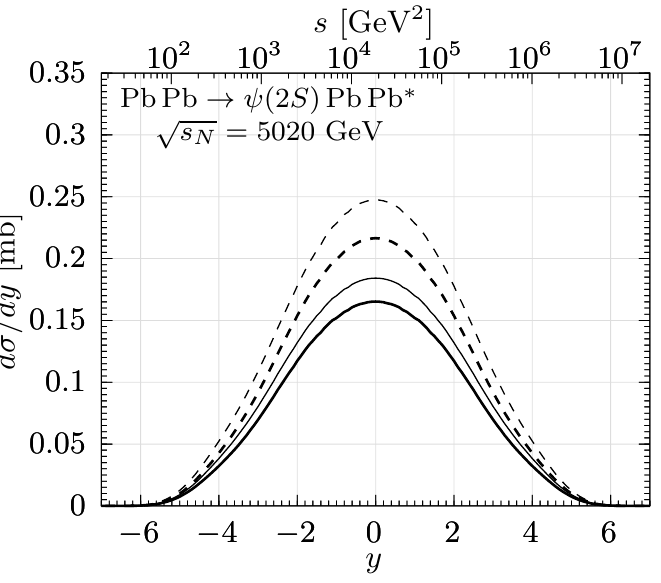}~~~~\\
\vspace*{-0.30cm}
\Caption{
  \label{Fig-UPC1psi2S}
  The same as Fig.~\ref{Fig-UPC1psi} but for the $\psip(2S)$ production in UPC at the
  LHC collision energy $\sqrt{s_N} = 2.76\,\TeV$ (top panels) and $\sqrt{s_N} = 5.02\,\TeV$ (bottom panels). The experimental value at $y=0$ has been obtained by the ALICE \cite{Adam:2015sia} collaboration.
  }
\EF
%%%%%%%%%%%%%%%%%%%%%%%%%%%%%%%%%%%%%%%%%%%%%%%%%%%%%%%%%%%%%%%%%%%%

%
%%%%%%%%%%%%%%%%%%%%%%%%%%%%%%%%%%%%%%%%%%%%%%%%%%%%%%%%%%%%%%%%%%%%
\BF
\hspace{-0.5cm}
\PSfig{1.1}{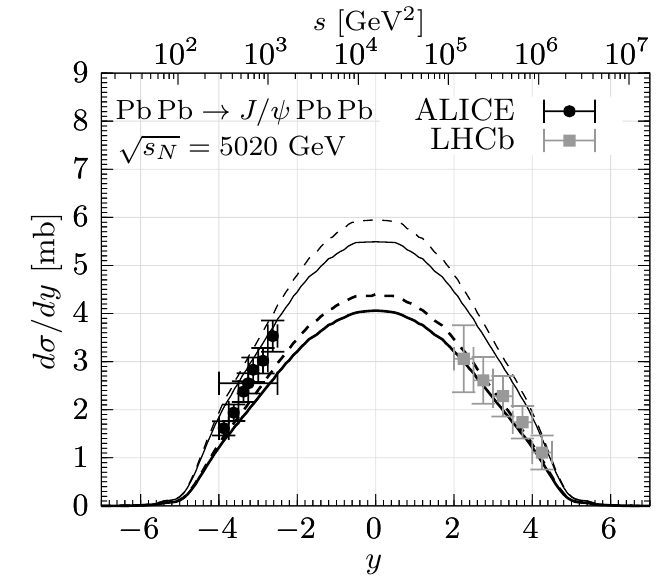}~~~~
\PSfig{1.1}{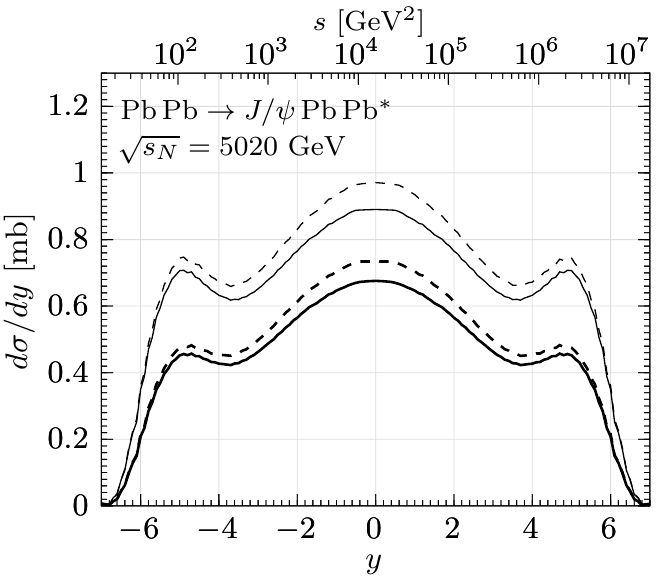}~~~~\\
\PSfig{1.1}{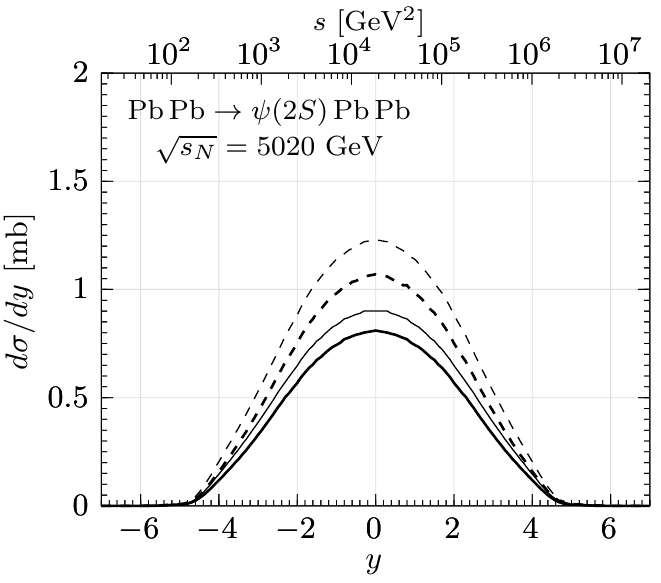}~~~~
\PSfig{1.1}{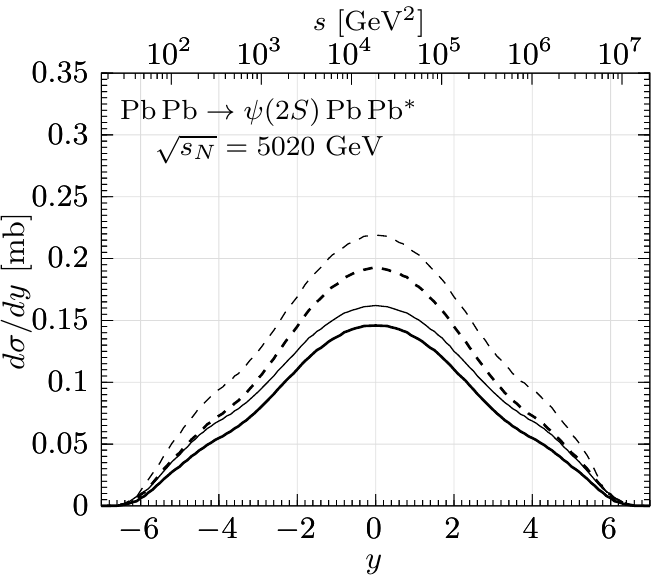}~~~~
\vspace{-0.3cm}
\Caption{
  \label{Fig-UPC1-IPsat}
  The same as Figs.~\ref{Fig-UPC1psi} and \ref{Fig-UPC1psi2S} but for the $\Jpsi(1S)$ (top panels) and $\psip(2S)$ (bottom panels) production in UPC at the LHC collision energy $\sqrt{s_N} = 5.02\,\TeV$. The predictions for $d\sigma/dy$ are based on the IPsat model \cite{Rezaeian:2012ji} for the dipole cross section.
  }
\EF
%%%%%%%%%%%%%%%%%%%%%%%%%%%%%%%%%%%%%%%%%%%%%%%%%%%%%%%%%%%%%%%%%%%%
%

One can see from Fig.~\ref{Fig-UPC1psi} that the inherence of $D$-wave component in charmonium wave functions, manifested itself as a difference between dashed and solid lines, is maximal at midrapidity ($y=0$) and causes the $7\%-10\%$ undesirable enhancement of $d\sigma/dy$ depending on the collision energy $\sqrt{s_N}$. Such a result is not affected by the shape of quarkonium wave functions generated by the BT and POW $c-\bar c$ interaction potentials used in our analysis.

%%%%%%%%%%%%%%%%%%%%%%%%%%%%%%%%%%%%%%%%%%%%%%%%%%%%%%%
Here we would like to emphasize that
a weak onset of $D$-wave effects demonstrated in Fig.~\ref{Fig-UPC1psi} is in accordance with 
our previous studies \cite{Krelina:2019egg}, as well as with a conclusion in Ref.~\cite{Lappi:2020ufv} about the lack of a significant contribution of $D$-wave component in $1S$ charmonium electroproduction. However,
such a contribution has not been quantified by means of a direct comparison of model predictions for charmonium electroproduction cross sections including and neglecting $D$-wave admixture. 
Moreover, the boosted Gaussian wave functions used in the analysis of relativistic corrections are not very suitable for description of charmonium states, especially radially excited states. They lead to antishadowing effects in the nucleus-to-nucleon ratios of photoproduction cross sections as was pointed out a long time ago in Refs.~\cite{Kopeliovich:1991pu,Kopeliovich:1993gk} for incoherent production of $\psip(2S)$.
However, such an enhanced nuclear production of $\psip(2S)$ with respect to a proton target is in contradiction with later studies (see Ref.~\cite{Ivanov:2002kc}, for example) based on realistic potential models for the $Q-\bar Q$ interaction. 
The reason is that in comparison with these models, the boosted Gaussian wave function for the $\psip(2S)$ state has a node with a position shifted to a smaller value of the dipole size as was analyzed in Ref~\cite{Cepila:2019skb}. Consequently, this enhances in the process $\gamma p\to \psip p$
the compensation from regions below and above the node position, i.e. the node effect becomes stronger.
The future experimental investigation of photoproduction of radially excited charmonium states off protons and nuclei will be very effective for ruling out various phenomenological models for quarkonium wave functions.
%%%%%%%%%%%%%%%%%%%%%%%%%%%%%%%%%%%%%%%%%%%%%%%%%%%%%%%

 %%%%%%%%%%%%%%%%%%%%%%%%%%%%%%%%%%%%%%%%%%%%%%%%%%%%%%%%%%%%%
\BT
\begin{tabular}{|l|c|c|c|c|}
 \hline
 ~~\boldmath{$Pb~Pb\to V~Pb~Pb$}~~
 & ~$\mathbf{dipole~cross~section}$~
 & ~$\mathbf{potential}$~
 & ~~$\mathbf{Scenario~ I}$~~~
 & ~~$\mathbf{Scenario~ II}$~~~\\
 \hline
 $~~V = \Jpsi(1S)$
 & GBW
 & POW 
 & $7.221~~\mb$ & $6.660~~\mb$\\
 $~~V = \psip(2S)$
 & GBW
 & POW
 & $1.530~~\mb$ & $1.134~~\mb$\\
 \hline
 \hline
\boldmath$~{\psip(2S)/\Jpsi(1S)}$
& $\mathbf{GBW}$ 
& $\mathbf{POW}$
& $\mathbf{0.212}$~~~~~~~ & $\mathbf{0.170}$~~~~~~~  \\
 \hline
 \hline
$~V = \Jpsi(1S)$
& GBW
& BT 
& $5.316~~\mb$ & $4.910~~\mb$\\
$~~V = \psip(2S)$
& GBW
& BT
& $1.300~~\mb$ & $0.989~~\mb$\\
\hline
\hline
\boldmath$~{\psip(2S)/\Jpsi(1S)}$
& $\mathbf{GBW}$
& $\mathbf{BT}$
& $\mathbf{0.245}$~~~~~~~ & $\mathbf{0.201}$~~~~~~~  \\
 \hline
 \hline
 $~~V = \Jpsi(1S)$
 & IPsat
 & POW 
 & $5.950~~\mb$ & $5.490~~\mb$\\
 $~~V = \psip(2S)$
 & IPsat
 & POW
 & $1.230~~\mb$ & $0.901~~\mb$\\
 \hline
 \hline
\boldmath$~{\psip(2S)/\Jpsi(1S)}$
& $\mathbf{IPsat}$
& $\mathbf{POW}$
& $\mathbf{0.207}$~~~~~~~ & $\mathbf{0.164}$~~~~~~~  \\
 \hline
 \hline
$~~V = \Jpsi(1S)$
& IPsat
& BT 
& $4.400~~\mb$ & $4.062~~\mb$\\
$~~V = \psip(2S)$
& IPsat
& BT
& $1.070~~\mb$ & $0.810~~\mb$\\
\hline
\hline
\boldmath$~{\psip(2S)/\Jpsi(1S)}$
& $\mathbf{IPsat}$
& $\mathbf{BT}$
& $\mathbf{0.243}$~~~~~~~ & $\mathbf{0.199}$~~~~~~~  \\
 \hline
 %\hline
\end{tabular}
%\label{Tab-2s-1s-y0}
\Caption{
  Values of nuclear cross sections at midrapidity for coherent production of $\Jpsi(1S)$ and $\psip(2S)$ in UPC at the LHC collision energy $\sqrt{s_N} = 5.02\,\TeV$ and the corresponding
  $\psip(2S)$-to-$\Jpsi(1S)$ ratios. Model predictions have been performed for the GBW and IPsat phenomenological models for the dipole cross section and for the quarkonium wave functions generated by the BT and POW potentials.
}
\label{Tab-2s-1s-y0}
\ET
%%%%%%%%%%%%%%%%%%%%%%%%%%%%%%%%%%%%%%%%%%%%%%%%%%%%%%%%%%%% %

%
%%%%%%%%%%%%%%%%%%%%%%%%%%%%%%%%%%%%%%%%%%%%%%%%%%%%%%%%%%%%%%%%%%%%
\BF
\hspace*{-0.4cm}
\PSfig{1.1}{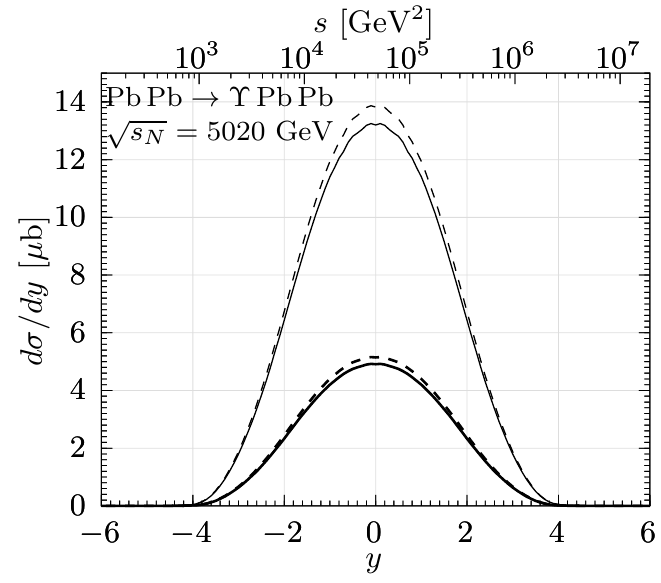}~~~~
\PSfig{1.1}{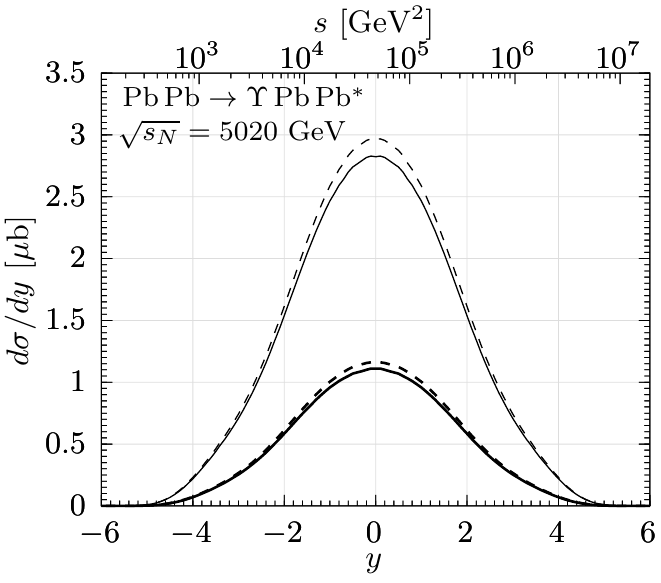}~~~~\\
\PSfig{1.1}{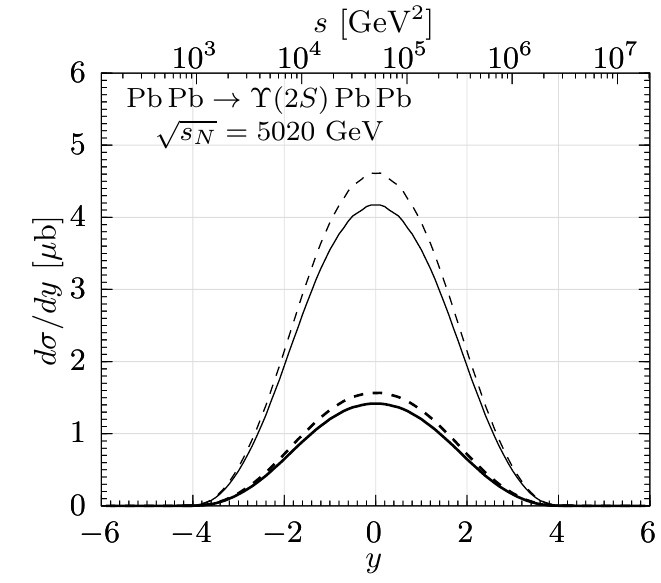}~~~~
\PSfig{1.1}{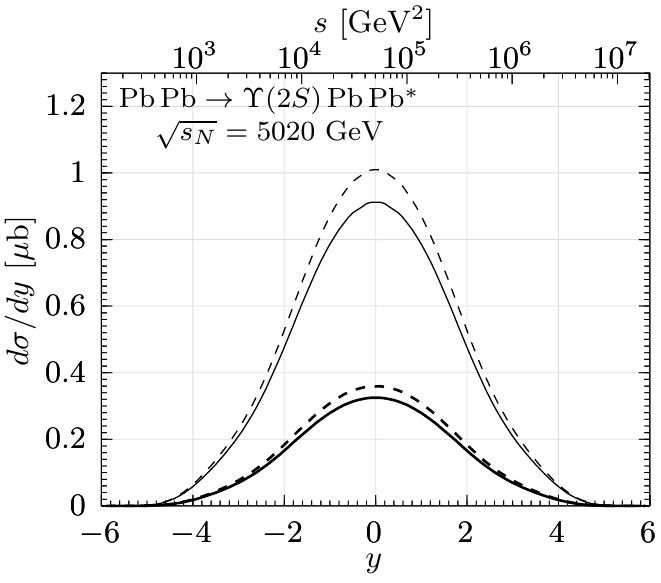}~~~~
\vspace{-0.3cm}
\Caption{
  \label{Fig-UPC1y}
  The same as Fig.~\ref{Fig-UPC1psi} but for the $\Y(1S)$ (top panels) and $\Yp(2S)$ (bottom panels) production in UPC at the LHC collision energy $\sqrt{s_N} = 5.02\,\TeV$. Here the bottomonium wave functions are generated by the BT potential. The thin and thick lines correspond to calculations using the GBW and KST models for the dipole cross section, respectively.
  }
\EF
%%%%%%%%%%%%%%%%%%%%%%%%%%%%%%%%%%%%%%%%%%%%%%%%%%%%%%%%%%%%%%%%%%%%
%

Figure~\ref{Fig-UPC1psi} also demonstrates that
a rather weak onset of $D$-wave effects 
can be hardly identified 
by future measurements of $\Jpsi(1S)$ production in UPC. 
However, there is a chance for recognition of such effects in $\psip(2S)$ production. Here the negative role of $D$-wave admixture is boosted due to a nodal structure of charmonium wave functions for excited states as is presented in Fig.~\ref{Fig-UPC1psi2S} for the $2S$ state. One can see that undesirable enhancement of $d\sigma/dy$ is now much larger than for the $1S$ state and represents $\sim 30\%-35\%$ in the LHC energy range. 
We have found that this ballast modification of $d\sigma/dy$ is still higher for the $\psipp(3S)$ state, due to two-node structure of the corresponding wave function, and reaches almost $50\%$.
Consequently, such a spurious $D$-wave manifestation is much stronger than other theoretical uncertainties related to different phenomenological models for $\sqq$,
as well as to different shapes of charmonium wave functions generated by various realistic $c-\bar c$ interaction potentials. More precise future measurements can help to identify and subsequently to eliminate such $D$-wave component and thus can be effective for the study of the quarkonium vertex structure.

%%%%%%%%%%%%%%%%%%%%%%%%%%%%%%%%%%%%%%%%%%%%%%%%%%%%%%%%%%%%%%%%%%%
In order to demonstrate that the relative $D$-wave contribution is practically independent of the model for the dipole cross section, we show in Fig.~\ref{Fig-UPC1-IPsat} our predictions for $d\sigma/dy$ using a more recent IPsat dipole model \cite{Rezaeian:2012ji}. Here a relative enhancement of $d\sigma/dy$ due to $D$-wave effects represents $\sim 8\%-9\,\%$ for 
coherent (left panel) and incoherent (right panel) production of $1S$ charmonia (top panels)
in UPC at the LHC collision energy $\sqrt{s_N}=5.02\,\TeV$. However, it reaches much higher value
$\sim 35\,\%$ for production of the $2S$ charmonium state (bottom panels). Such modifications of $d\sigma/dy$ correspond to the above results using the GBW dipole parametrization (see Figs.~\ref{Fig-UPC1psi} and \ref{Fig-UPC1psi2S}).
%%%%%%%%%%%%%%%%%%%%%%%%%%%%%%%%%%%%%%%%%%%%%%%%%%%%%%%%%%%%%%%%%%%

%============================================================================================
The effective and alternative way for the identification of undesirable $D$-wave effects, related to the photonlike structure of the quarkonium vertex, is based on the study of the $\psip(2S)$-to-$\Jpsi(1S)$ ratio of nuclear cross sections. This allows to eliminate to a certain extent theoretical uncertainties inherent in calculations of rapidity distributions $d\sigma/dy$ of coherent as well as incoherent charmonium photoproduction within the LF color dipole formalism. Since new ALICE data on coherent production of $\Jpsi(1S)$ and $\psip(2S)$ at midrapidities will appear soon, in Table~\ref{Tab-2s-1s-y0} we present our predictions for nuclear cross sections and for the corresponding $\psip(2S)$-to-$\Jpsi(1S)$ ratio $R_{\psip/\Jpsi}(y) = \frac{d\sigma_A^{coh}(\psip)/dy}{d\sigma_A^{coh}(\Jpsi)/dy}$ at $y=0$. Here we 
adopt the GBW and IPsat dipole models, as well as quarkonium wave functions determined by the BT and POW potential. One can see that the \textit{scenario I} gives the ratio $R_{\psip/\Jpsi}(y=0)\sim 0.207\div 0.245$, which differs 
%sufficiently 
from values $\sim 0.164\div 0.201$ corresponding to \textit{scenario II}. Consequently, 
such a difference should be sufficient for a recognition between both scenarios based 
on a comparison with the forthcoming more precise ALICE data.
%will allow us to make a final and definite conclusion about the preference of the \textit{Scenario I} or the \textit{Scenario II}.
%===========================================================================================

Figure~\ref{Fig-UPC1y} clearly shows a weak negative role of $D$-wave component
in $1S$ (top panels) and $2S$ (bottom panels) bottomonium production in UPC at $\sqrt{s_N} = 5.02\,\TeV$.
It causes only $\sim 5\%$ modification of $d\sigma/dy$ for $\Y(1S)$ production (see differences between dashed and solid lines in top panels). Similarly as for the $\psip(2S)$ state,
the nodal structure of the wave function leads to a stronger onset of $D$-wave effects in $\Yp(2S)$ production in UPC compared to the $\Y(1S)$ state. This
is demonstrated in the bottom panels of Fig.~\ref{Fig-UPC1y} where the photonlike structure of $\Yp(2S)\to b\bar b$ transition causes $\sim 10\%-12\%$ enhancement of $d\sigma/dy$ with respect to the ``$S$-wave-only'' \textit{scenario II}. We have also estimated that
for production of the $\Ypp(3S)$ state this enhancement is still stronger, reaching $\sim 15\%$.
However, such undesirable modifications are still 
smaller than other theoretical uncertainties originated from different models for $\sqq$
(see thin and thick lines in Fig.~\ref{Fig-UPC1y}).
For this reason, the production of bottomonia in UPC is not effective for the study of the structure of quarkonium $V\to Q\bar Q$ transition.

%%%%%%%%%%%%%%%%%%%%%%%%%%%%%%%%%
\section{Conclusions}
\label{Sec:3}
%%%%%%%%%%%%%%%%%%%%%%%%%%%%%%%%%

In this paper, we have analyzed how the $D$-wave admixture in quarkonium wave functions, related to the photonlike structure of $V\to Q\bar Q$ transition,
can falsify the model predictions for distributions $d\sigma/dy$ in production of various heavy quarkonia in heavy-ion UPC. Our calculations are based on the standard formulas for nuclear cross sections within the LF color dipole approach. The model results include also nuclear phenomena related to higher twist quark shadowing, as well as the leading twist gluon shadowing. 

The onset of the $D$-wave component was quantified by comparing two scenarios. \textit{Scenario I} corresponds to the photonlike structure of the $V\to Q\bar Q$ transition. Such a structure is imposed in the LF frame and is treated in most of the recent papers.
It leads to the $D$-wave component presented in various model predictions with its corresponding role, which is not justified by any nonrelativistic $Q-\bar Q$ interaction potential.
\textit{Scenario II} is based on a simple ``$S$-wave-only'' structure of quarkonium wave functions in the $Q\bar Q$ rest frame.

We have found that all calculations based on \textit{scenario I}
enhance the magnitude of $d\sigma/dy$ 
compared to \textit{scenario II}, especially at $y=0$. Our calculations confirm that the onset of $D$-wave effects is rather weak in production of $\Jpsi(1S)$, $\Y(1S)$ and $\Yp(2S)$ states where the corresponding undesirable modification of nuclear cross section does not exceed $\sim 10\%-12\%$ and is smaller than other theoretical uncertainties related mainly to the shape of quarkonium wave functions, as well as to models for the dipole cross section.

However, there is a chance to identify and eventually to eliminate a negative role of $D$-wave effects in theoretical predictions 
and so to abandon the unjustified assumption about the photonlike structure of quarkonium vertex.
According to this, we propose to
investigate the production of higher charmonium states in UPC, such as $\psip(2S)$, $\psipp(3S)$, etc. In this case, a spurious enhancement of nuclear cross sections at midrapidities due to \textit{scenario I} exceeds at least $35\%$ compared to ``$S$-wave-only'' \textit{scenario II}
and is larger than other theoretical uncertainties. 
Besides, such uncertainties can be reduced by studying production of the $\psip(2S)$-to-$\Jpsi(1S)$ ratio.
This gives a possibility that more precise photoproduction data at $y=0$ from the future measurements at the LHC, as well as at planned electron-ion colliders can shed more light on the structure of $V\to Q\bar Q$ transition.

%%%%%%%%%%%%%%%%%%%%%%%%%%%%%%%%%
\noindent\textbf{Acknowledgment}: 
%%%%%%%%%%%%%%%%%%%%%%%%%%%%%%%%%
This work was supported in part by the project of the
European Regional Development Fund No. CZ.02.1.01/0.0/0.0/16\_019/0000778.
The work of J.N. was partially supported by Grant
No. LTT18002 of the Ministry of Education, Youth and
Sports of the Czech Republic and by the Slovak Funding
Agency, Grant No. 2/0007/18.

%J.N. work was partially supported by the grant 
%LTT18002 of the Ministry of  Education,  Youth  and  Sports  of  the  Czech  Republic,  by  the project of the European Regional Development Fund CZ02.1.01/0.0/0.0/16\_019/0000778 and by the Slovak Funding Agency, Grant 2/0007/18.
%
%The work of M.K. was supported by the project Centre of Advanced Applied Sciences with the number: CZ.02.1.01/0.0/0.0/16-019/0000778 (Czech Republic). Project Centre of Advanced Applied Sciences is co-financed by European Union.

% ===========================================================

% -----------------------------------------------------------


\begin{thebibliography}{99}
% -----------------------------------------------------------

%%%%%%%%%%%%%%%%%%%%%%%%%%%%%%%%%%%%%%%%%%%%%%% 1 %%%%%%%%%%%%%%%%%%%%%%%%%%%%%%%%%%%%%

\bibitem{Kopeliovich:1991pu} 
  B.Z.~Kopeliovich and B.G.~Zakharov;
  %``Quantum effects and color transparency in charmonium photoproduction on nuclei,''
  Phys.\ Rev.\ D \textbf{44}, 3466 (1991).

\bibitem{Kopeliovich:1993pw} 
  B.Z.~Kopeliovich, J.~Nemchik, N.N.~Nikolaev and B.G.~Zakharov;
  %``Decisive test of color transparency in exclusive electroproduction of vector mesons,''
  Phys.\ Lett.\ B \textbf{324}, 469 (1994).

\bibitem{Nemchik:1996pp}
  J.~Nemchik, N.~N.~Nikolaev, E.~Predazzi and B.~Zakharov,
  %``Color dipole systematics of diffractive photoproduction and electroproduction of vector mesons,''
  Phys.\ Lett.\ B \textbf{374}, 199 (1996).
  %doi:10.1016/0370-2693(96)00153-0
  %[arXiv:hep-ph/9604419 [hep-ph]].
  %71 citations counted in INSPIRE as of 06 Apr 2020

\bibitem{Nemchik:1996cw}
  J.~Nemchik, N.~N.~Nikolaev, E.~Predazzi and B.~Zakharov,
  %``Color dipole phenomenology of diffractive electroproduction of light vector mesons at HERA,''
  Z.\ Phys.\ C \textbf{75}, 71 (1997).
  %doi:10.1007/s002880050448
  %[arXiv:hep-ph/9605231 [hep-ph]].
  %190 citations counted in INSPIRE as of 06 Apr 2020

\bibitem{Nemchik:2002ug}
  J.~Nemchik,
  %``Incoherent production of charmonia off nuclei as a good tool for study of color transparency,''
  Phys.\ Rev.\ C \textbf{66}, 045204 (2002).
  %doi:10.1103/PhysRevC.66.045204
  %[arXiv:hep-ph/0205276 [hep-ph]].
  %6 citations counted in INSPIRE as of 06 Apr 2020


%%%%%%%%%%%%%%%%%%%%%%%%%%%%%%%%%%%%%%%%%%%% 2 %%%%%%%%%%%%%%%%%%%%%%%%%%%%%%%%%%%%%%%%%%%%%%

\bibitem{Krelina:2018hmt}
  M.~Krelina, J.~Nemchik, R.~Pasechnik and J.~Cepila,
  %``Spin rotation effects in diffractive electroproduction of heavy quarkonia,''
  Eur.\ Phys.\ J.\ C \textbf{79}, 154 (2019).
  %doi:10.1140/epjc/s10052-019-6666-y
  %[arXiv:1812.03001 [hep-ph]].
  %5 citations counted in INSPIRE as of 06 Apr 2020

\bibitem{Cepila:2019skb}
  J.~Cepila, J.~Nemchik, M.~Krelina and R.~Pasechnik,
  %``Theoretical uncertainties in exclusive electroproduction of S-wave heavy quarkonia,''
  Eur.\ Phys.\ J.\ C \textbf{79}, 495 (2019).
  %doi:10.1140/epjc/s10052-019-7016-9
  %[arXiv:1901.02664 [hep-ph]].
  %10 citations counted in INSPIRE as of 06 Apr 2020

\bibitem{Krelina:2019egg}
  M.~Krelina, J.~Nemchik and R.~Pasechnik,
  %``$D$-wave effects in diffractive electroproduction of heavy quarkonia from the photon-like $V\rightarrow Q\bar Q$ transition,''
  Eur.\ Phys.\ J.\ C \textbf{80}, 92 (2020).
  %doi:10.1140/epjc/s10052-020-7678-3
  %[arXiv:1909.12770 [hep-ph]].
  %2 citations counted in INSPIRE as of 06 Apr 2020

\bibitem{Kopeliovich:2020has}
  B.~Z.~Kopeliovich, M.~Krelina, J.~Nemchik and I.~K.~Potashnikova,
  %``Heavy quarkonium production in ultraperipheral nuclear collisions,''
  arXiv:2008.05116.
  %0 citations counted in INSPIRE as of 19 Aug 2020

\bibitem{Kopeliovich:2001xj}
  B.~Kopeliovich, J.~Nemchik, A.~Schafer and A.~Tarasov,
  %``Color transparency versus quantum coherence in electroproduction of vector mesons off nuclei,''
  Phys.\ Rev.\ C \textbf{65}, 035201 (2002).
  %doi:10.1103/PhysRevC.65.035201
  %[arXiv:hep-ph/0107227 [hep-ph]].
  %104 citations counted in INSPIRE as of 06 Apr 2020

%%%%%%%%%%%%%%%%%%%%%%%%%%%%%%%%%%%%%%%%% 3 %%%%%%%%%%%%%%%%%%%%%%%%%%%%%%%%%%%%%%%%%%%%%%

\bibitem{Kopeliovich:2000ra} 
  B.Z.~Kopeliovich, J.~Raufeisen and A.V.~Tarasov,
  %``Nuclear shadowing and coherence length for longitudinal and transverse photons,''
  Phys.\ Rev.\ C \textbf{62}, 035204 (2000).
  %doi:10.1103/PhysRevC.62.035204
  %[hep-ph/0003136].
  %%CITATION = doi:10.1103/PhysRevC.62.035204;%%
  %87 citations counted in INSPIRE as of 31 Jan 2020

\bibitem{Kopeliovich:1999am} 
  B.Z.~Kopeliovich, A.~Schafer and A.V.~Tarasov;
  %``Nonperturbative effects in gluon radiation and photoproduction of quark pairs,''
  Phys.\ Rev.\ D \textbf{62}, 054022 (2000).


\bibitem{Krelina:2020ipn}
M.~Krelina and J.~Nemchik,
%``Nuclear shadowing in DIS at electron-ion colliders,''
Eur. Phys. J. Plus \textbf{135}, 444 (2020).
%doi:10.1140/epjp/s13360-020-00498-2
%[arXiv:2003.04156 [hep-ph]].
%1 citations counted in INSPIRE as of 08 Sep 2020


\bibitem{Kopeliovich:1993gk}
  B.~Kopeliovich, J.~Nemchik, N.~N.~Nikolaev and B.~Zakharov,
  %``Novel color transparency effect: Scanning the wave function of vector mesons,''
  Phys.\ Lett.\ B \textbf{309}, 179 (1993).
  %doi:10.1016/0370-2693(93)91523-P
  %[arXiv:hep-ph/9305225 [hep-ph]].
  %138 citations counted in INSPIRE as of 06 Apr 2020

\bibitem{Nemchik:1994fp}
  J.~Nemchik, N.~N.~Nikolaev and B.~Zakharov,
  %``Scanning the BFKL pomeron in elastic production of vector mesons at HERA,''
  Phys.\ Lett.\ B \textbf{341}, 228 (1994).
  %doi:10.1016/0370-2693(94)90314-X
  %[arXiv:hep-ph/9405355 [hep-ph]].
  %167 citations counted in INSPIRE as of 06 Apr 2020

%%%%%%%%%%%%%%%%%%%%%%%%%%%%%%%%%%%%%%%% 4 %%%%%%%%%%%%%%%%%%%%%%%%%%%%%%%%%%%%%%%%%%%%%%

\bibitem{Kopeliovich:2007wx}
  B.~Kopeliovich, J.~Nemchik and I.~Schmidt,
  %``Production of Polarized Vector Mesons off Nuclei,''
  Phys.\ Rev.\ C \textbf{76}, 025210 (2007).
  %doi:10.1103/PhysRevC.76.025210
  %[arXiv:hep-ph/0703118 [hep-ph]].
  %8 citations counted in INSPIRE as of 06 Apr 2020

\bibitem{Nemchik:1994fq}
  J.~Nemchik, N.~N.~Nikolaev and B.~Zakharov,
  %``Anomalous a-dependence of diffractive leptoproduction of radial excitation rho-prime (2s),''
  Phys.\ Lett.\ B \textbf{339}, 194 (1994).
  %doi:10.1016/0370-2693(94)91154-1
  %[arXiv:nucl-th/9405025 [nucl-th]].
  %27 citations counted in INSPIRE as of 06 Apr 2020

\bibitem{Hufner:2000jb} 
  J.~Hufner, Y.P.~Ivanov, B.Z.~Kopeliovich and A.V.~Tarasov;
  %``Photoproduction of charmonia and total charmonium proton cross-sections,''
  Phys.\ Rev.\ D \textbf{62}, 094022 (2000).

\bibitem{jan-00a} 
   J. Nemchik; 	
   %\textit{Wave function of 2S radially excited vector mesons 
   %from data for diffraction slope},
   Phys. Rev. D \textbf{63}, 074007 (2001).

\bibitem{jan-00b} 
   J. Nemchik; 	
   %\textit{Anomalous t dependence in diffractive electroproduction 
   %of 2S radially excited light vector mesons at HERA},
   Eur. Phys. J. C \textbf{18}, 711 (2001).

%%%%%%%%%%%%%%%%%%%%%%%%%%%%%%%%%%%%%%% 5 %%%%%%%%%%%%%%%%%%%%%%%%%%%%%%%%%%%%%%%%%%%%%%%

\bibitem{Buchmuller:1980su} 
  W.~Buchmuller and S.H.H.~Tye;
  %``Quarkonia and Quantum Chromodynamics,''
  Phys.\ Rev.\ D \textbf{24}, 132 (1981).  

\bibitem{Martin:1980jx}
  A.~Martin;
  %``A FIT of Upsilon and Charmonium Spectra,''
  Phys.\ Lett.\ \textbf{93B}, 338 (1980).

\bibitem{Barik:1980ai}
  N.~Barik and S.N.~Jena;
  %``Fine - Hyperfine Splittings Of Quarkonium Levels In An Effective Power Law Potential,''
  Phys.\ Lett.\ \textbf{97B}, 265 (1980). 
  
\bibitem{GolecBiernat:1998js} 
  K.J.~Golec-Biernat and M.~Wusthoff;
  %``Saturation effects in deep inelastic scattering at low Q**2 and its implications on diffraction,''
  Phys.\ Rev.\ D \textbf{59}, 014017 (1998).
  
\bibitem{GolecBiernat:1999qd} 
  K.J.~Golec-Biernat and M.~Wusthoff;
  %``Saturation in diffractive deep inelastic scattering,''
  Phys.\ Rev.\ D \textbf{60}, 114023 (1999).

%%%%%%%%%%%%%%%%%%%%%%%%%%%%%%%%%%%%%%%%%% 6 %%%%%%%%%%%%%%%%%%%%%%%%%%%%%%%%%%%%%%%%%%%%


\bibitem{Terentev:1976jk} 
  M.V.~Terentev;
  %``On the Structure of Wave Functions of Mesons as Bound States of Relativistic Quarks,''
  Sov.\ J.\ Nucl.\ Phys.\ \textbf{24}, 106 (1976)
  [Yad.\ Fiz.\ \textbf{24}, 207 (1976)].

\bibitem{Melosh:1974cu} 
  H.J.~Melosh;
  %``Quarks: Currents and constituents,''
  Phys.\ Rev.\ D \textbf{9}, 1095 (1974).



\bibitem{Ivanov:2002kc}
  Y.~Ivanov, B.~Kopeliovich, A.~Tarasov and J.~Hufner,
  %``Electroproduction of charmonia off nuclei,''
  Phys.\ Rev.\ C \textbf{66}, 024903 (2002).
  %doi:10.1103/PhysRevC.66.024903
  %[arXiv:hep-ph/0202216 [hep-ph]].
  %29 citations counted in INSPIRE as of 06 Apr 2020

\bibitem{DeJager:1987qc}
  H.De Vries, C.W.De Jager and C.De Vries,
  %``Nuclear charge and magnetization density distribution parameters from elastic electron scattering,''
  At.\ Data Nucl.\ Data Tables\ \textbf{36}, 495 (1987).
%  doi:10.1016/0092-640X(87)90013-1

\bibitem{zkl} 
   B.~Z.~Kopeliovich, L.~I.~Lapidus and A.~B.~Zamolodchikov,
   %``Dynamics of Color in Hadron Diffraction on Nuclei,''
   JETP Lett. \textbf{33}, 595 (1981).
   %JINR-E2-81-251.

%%%%%%%%%%%%%%%%%%%%%%%%%%%%%%%%%%%%%%%% 7 %%%%%%%%%%%%%%%%%%%%%%%%%%%%%%%%%%%%%%%%%%%%%%

\bibitem{Ivanov:2002eq}
Y.~P.~Ivanov, B.~Kopeliovich, A.~Tarasov and J.~Hufner,
  %``Electroproduction of charmonia off protons and nuclei,''
  AIP Conf. Proc. \textbf{660}, 283 (2003).
  %AIP Conf. Proc. \textbf{660}, no.1, 283 (2003).
  %doi:10.1063/1.1570580
  %[arXiv:hep-ph/0212322 [hep-ph]].
  %5 citations counted in INSPIRE as of 06 May 2020

\bibitem{Ivanov:2007ms}
  Y.~Ivanov, B.~Kopeliovich and I.~Schmidt,
  %``Vector meson production in ultra-peripheral collisions at LHC,''
  arXiv:0706.1532.
  %23 citations counted in INSPIRE as of 06 May 2020

\bibitem{bronzan-74}
   J.B. Bronzan, G.L. Kane and U.P. Sukhatme;
   Phys. Lett. \textbf{49B}, 272 (1974).

\bibitem{forshaw-03} 
   J.R. Forshaw, R. Sandapen and G. Shaw;
   %``Color dipoles and rho, phi electroproduction,''
   Phys. Rev. D \textbf{69}, 094013 (2004).

\bibitem{Nikolaev:1994ce}
  N.~N.~Nikolaev and B.~Zakharov,
  %``On determination of the large 1/x gluon distribution at HERA,''
  Phys. Lett. B \textbf{332}, 184 (1994).
  %doi:10.1016/0370-2693(94)90877-X
  %[arXiv:hep-ph/9403243 [hep-ph]].
  %153 citations counted in INSPIRE as of 07 May 2020

%%%%%%%%%%%%%%%%%%%%%%%%%%%%%%%%%%%%%%%%%% 8 %%%%%%%%%%%%%%%%%%%%%%%%%%%%%%%%%%%%%%%%%%%%%

\bibitem{jan-98}
   J.~Nemchik, N.N.~Nikolaev, E.~Predazzi, B.G.~Zakharov and V.R.~Zoller;
   %\textit{The diffration cone for exclusive vector meson production
   %        in deep inelastic scattering}
   J. Exp. Theor. Phys. \textbf{86}, 1054 (1998).

\bibitem{Kutak:2003bd}
   K.~Kutak and J.~Kwiecinski,
   %``Screening effects in the ultrahigh-energy neutrino interactions,''
   Eur. Phys. J. C \textbf{29}, 521 (2003).
   %doi:10.1140/epjc/s2003-01236-y
   %[arXiv:hep-ph/0303209 [hep-ph]].
   %109 citations counted in INSPIRE as of 11 Jun 2020

%
% ...... Data on charmonium production in UPCs ...................................
%

% ...... coherent J/\Psi .............. 200 GeV

%\bibitem{Afanasiev:2009hy}
%S.~Afanasiev \textit{et al.} [PHENIX],
%``Photoproduction of J/psi and of high mass e+e- in ultra-peripheral Au+Au collisions at %s**(1/2) = 200-GeV,''
%  Phys. Lett. B \textbf{679}, 321-329 (2009).
  %doi:10.1016/j.physletb.2009.07.061
  %[arXiv:0903.2041 [nucl-ex]].
%108 citations counted in INSPIRE as of 06 May 2020

% ...... coherent J/\Psi .............. 2.76 TeV

\bibitem{Khachatryan:2016qhq}
  V.~Khachatryan \textit{et al.} (CMS Collaboration),
  %``Coherent $J/\psi$ photoproduction in ultra-peripheral PbPb collisions at $\sqrt {s_{NN}} =$ 2.76 TeV with the CMS experiment,''
  Phys. Lett. B \textbf{772}, 489-511 (2017).
  %doi:10.1016/j.physletb.2017.07.001
  %[arXiv:1605.06966 [nucl-ex]].
  %69 citations counted in INSPIRE as of 17 Apr 2020

\bibitem{Abelev:2012ba}
  B.~Abelev \textit{et al.} (ALICE Collaboration),
  %``Coherent $J/\psi$ photoproduction in ultra-peripheral Pb-Pb collisions at $\sqrt{s_{NN}} = 2.76$ TeV,''
  Phys. Lett. B \textbf{718}, 1273-1283 (2013).
  %doi:10.1016/j.physletb.2012.11.059
  %[arXiv:1209.3715 [nucl-ex]].
  %220 citations counted in INSPIRE as of 17 Apr 2020
  
% ...... coherent + incoherent J/\Psi .............. 2.76 TeV

\bibitem{Abbas:2013oua}
  E.~Abbas \textit{et al.} (ALICE Collaboration),
  %``Charmonium and $e^+e^-$ pair photoproduction at mid-rapidity in ultra-peripheral Pb-Pb collisions at $\sqrt{s_{\rm NN}}$=2.76 TeV,''
  Eur. Phys. J. C \textbf{73}, no.11, 2617 (2013).
  %doi:10.1140/epjc/s10052-013-2617-1
  %[arXiv:1305.1467 [nucl-ex]].
  %209 citations counted in INSPIRE as of 17 Apr 2020

%%%%%%%%%%%%%%%%%%%%%%%%%%%%%%%%%%%%%%%%% 9 %%%%%%%%%%%%%%%%%%%%%%%%%%%%%%%%%%%%%%%%%%%%%%%

% ...... coherent \Psi' .............. 2.76 TeV

\bibitem{Adam:2015sia}
  J.~Adam \textit{et al.} (ALICE),
  %``Coherent $\psi$(2S) photo-production in ultra-peripheral Pb Pb collisions at $\sqrt{s}_{\rm NN}$ = 2.76 TeV,''
  Phys. Lett. B \textbf{751}, 358-370 (2015).
  %doi:10.1016/j.physletb.2015.10.040
  %[arXiv:1508.05076 [nucl-ex]].
  %38 citations counted in INSPIRE as of 17 Apr 2020

% ...... coherent J/\Psi .............. 5.02 TeV

\bibitem{LHCb:2018ofh}
  A.~Bursche (LHCb Collaboration),
  %``Study of coherent $J/\psi$ production in lead-lead collisions at $\sqrt{s_{\rm NN}} =5\ \rm{TeV}$ with the LHCb experiment,''
  Nucl. Phys. A \textbf{982}, 247-250 (2019).
  %doi:10.1016/j.nuclphysa.2018.10.069
  %19 citations counted in INSPIRE as of 17 Apr 2020

\bibitem{Acharya:2019vlb}
  S.~Acharya \textit{et al.} (ALICE Collaboration),
  %``Coherent J/$\psi$ photoproduction at forward rapidity in ultra-peripheral Pb-Pb collisions at $\sqrt{s_{\rm{NN}}}=5.02$ TeV,''
  Phys. Lett. B \textbf{798}, 134926 (2019).
  %doi:10.1016/j.physletb.2019.134926
  %[arXiv:1904.06272 [nucl-ex]].
  %11 citations counted in INSPIRE as of 17 Apr 2020

\bibitem{Lappi:2020ufv}
   T.~Lappi, H.~M\"antysaari and J.~Penttala,
   %``Relativistic corrections to the vector meson light front wave function,''
   Phys. Rev. D \textbf{102}, 054020 (2020)
   %doi:10.1103/PhysRevD.102.054020
   %[arXiv:2006.02830 [hep-ph]].
   %3 citations counted in INSPIRE as of 29 Oct 2020

\bibitem{Rezaeian:2012ji}
   A.H.~Rezaeian, M.~Siddikov, M.~Van de Klundert and R.~Venugopalan;
   %``Analysis of combined HERA data in the Impact-Parameter dependent Saturation model,''
   Phys. Rev. D\textbf{87} 034002 (2013).
   %doi:10.1103/PhysRevD.87.034002
   %[arXiv:1212.2974 [hep-ph]].


\end{thebibliography}
\end{document}